\documentclass[aps,prl,twocolumn,superscriptaddress,showpacs,draft]{revtex4}
\input{epsf}

\newcommand{\rem}[1]{}

\newcommand{\ps}{\ps}

\begin{document}

\title{Scarring in open quantum systems}

\author{Diego Wisniacki}
\affiliation{Departamento de F\'\i sica, FCEyN, UBA, Pabell\'on 1 Ciudad 
Universitaria, C1428EGA Buenos Aires, Argentina}

\author{Gabriel G. Carlo} 
\affiliation{Departamento de F\'\i sica, Comisi\'on Nacional de Energ\'\i a 
At\'omica, Avenida del Libertador 8250, (C1429BNP) Buenos Aires, Argentina}

\date{\today}


\begin{abstract}
We study scarring phenomena in open quantum systems. 
We show numerical evidence that individual resonance 
eigenstates of an open quantum system 
present localization around unstable short periodic 
orbits in a similar way as their closed counterparts. The structure of eigenfunctions 
around these classical objects is not destroyed 
by the opening. This is exposed in a paradigmatic 
system of quantum chaos, the cat map. 
\end{abstract}

\maketitle

Open quantum systems are very important in many areas of physics. 
For example, they play a central role in the study 
of quantum to classical correspondence \cite{Quantum2Classical}, 
microlasers \cite{Microlasers,WiersigPrl06}, 
quantum dots \cite{QuantumDots}, chaotic scattering 
\cite{ChaoticScattering}, and more. 
However, there are several properties of these systems that are 
less known if compared to those of closed ones. 

Quantum evolution in open systems is given 
by nonunitary matrices, whose eigenstates (resonances) are nonorthogonal and 
the eigenvalues are complex with modulus less than or equal to one.
One of the main conjectures about the properties 
of the spectrum is that the mean 
density of resonances follows the fractal Weyl law \cite{LuPrl03}. 
This law predicts that the number of eigenstates that have a 
finite decay rate goes as $N_\gamma \propto \hbar^{-(d-1)}$, where $d$ 
is a fractal dimension of the classical strange repeller.
This result has been tested in some systems 
\cite{SchomerusPrl04,ShepelyanskyarXiv07}.
As a consequence, the majority of the eigenfunctions become 
degenerate with their eigenvalue modulus tending to zero as the 
size of the opening (the number of decay channels) 
relative to $\hbar$ increases. These are the short lived eigenstates, 
which cannot be associated to any classical trapped set 
(instead, they can be related to the 
trajectories that escape from the system before the Ehrenfest time).    
On the other hand, the number of remaining eigenstates 
(the long-lived ones) tends to zero.
However, they contain the most relevant classical information, resembling the 
classical repeller. This was noticed in \cite{ShepelyanskyPhD99}, where 
they were coined quantum fractal eigenstates. Moreover,  
this investigation was recently extended \cite{KeatingPrl06} by looking at 
the right and left resonances of the open baker's map. It was found that  
the probability density averaged for several right eigenstates 
is supported by the classical Cantor set (the repeller), 
showing self-similarity both in the $q$ and $p$ representation.
Finally, in the more specific context of optical microcavities, the formation 
of long-lived scarred modes has been observed \cite{WiersigPrl06}. This 
behaviour has been associated to avoided resonance crossings.
Nevertheless, almost nothing else is known about the morphology 
of individual resonances.

We are interested in the study of quantum systems which are classically chaotic. 
In closed quantum chaotic systems, the morphology of the eigenfunctions 
has been extensively studied. One of the most important 
and striking properties is scarring \cite{HellerPrl84}. 
This consists of the localization, i.e., the probability enhancement of given 
individual eigenfunctions along short unstable periodic orbits (POs). 
This effect has been discovered in the Bunimovich stadium billiard \cite{McDonald} 
and a great amount of work has been done since then \cite{ScarTheory}, giving 
rise to what is known as ``scar theory''. 

In this letter we explore quantitatively the localization 
properties of resonances. We have studied the overlaps 
of wavefunctions highly localized on the vicinity of POs (scar functions) 
\cite{ScarFunction1,ScarFunction2,ScarFunction3} with the eigenfunctions 
of an open quantum system in order to unveil the quantum mechanical 
manifestation of short POs. 
These values result to be higher than when the overlap is calculated with 
the eigenfunctions of the closed system. The $\hbar$ smoothed fractal 
nature does not destroy structures 
of this kind. This effect is even greater 
when the area of the opening grows, thus it cannot be ascribed 
to a perturbative origin. We provide with an interpretation of these 
results.

%----------------------------------------------------------------------------
%results: model definitions, figures, description.

One of the most studied open systems correspond to two dimensional torus maps, 
where a band along the $q$ or $p$ directions is cut by means of a projection. 
The corresponding quantum dynamics is given by a nonunitary matrix 
$\bar{M}=P M$ (or equivalently $\bar{M}'=M P$ which is related to $\bar{M}$ by a 
time reversal operation), where $M$ is the closed map and $P$ is the projector on 
the complement of the opening. 
This quantum evolution is characterized by decaying eigenstates $\phi_i$, 
whose corresponding eigenvalues $z_i$ have complex energies.
It is usual to define $|z_i|^2=\exp{(-\Gamma_i)}$, where $\Gamma_i \geq 0$ 
is called the decay rate. We have studied a paradigmatic model 
of quantum chaos, the cat map, which 
is a linear automorphism on the 2-torus generated by the 2 x 2 symplectic 
matrix $\cal{M}$, modulus 1. We have used 
\begin{eqnarray}
\label{eq:ccat}
\cal{M}&=&\left(\begin{array}{cc}2&3\\1&2\end{array}\right). 
\end{eqnarray}
When quantizing the torus we have a finite Hilbert space of dimension 
$N=1/2\pi\hbar$ and a discrete $N$ lattice of position and momenta in the 
unit interval. The quantum cat map in the position representation 
is given by the matrix $M$ whose elements are \cite{HannayPhysD80}
\begin{equation}
\label{eq:qcat}
{M}_{kj}=\left( \frac{\rm i}{n} \right)^{1/2} 
\exp{\left[\frac{2\pi{\rm i}}{n}(k^2-jk+j^2)\right]}. 
\end{equation}
Finally, we choose to apply the projector $P$ after $M$ 
to obtain the nonunitary matrix $\bar M$ 
that gives the evolution of the open cat map.

\begin{figure}[htp] %[tp]
\centerline{\epsfxsize=7cm\epsffile{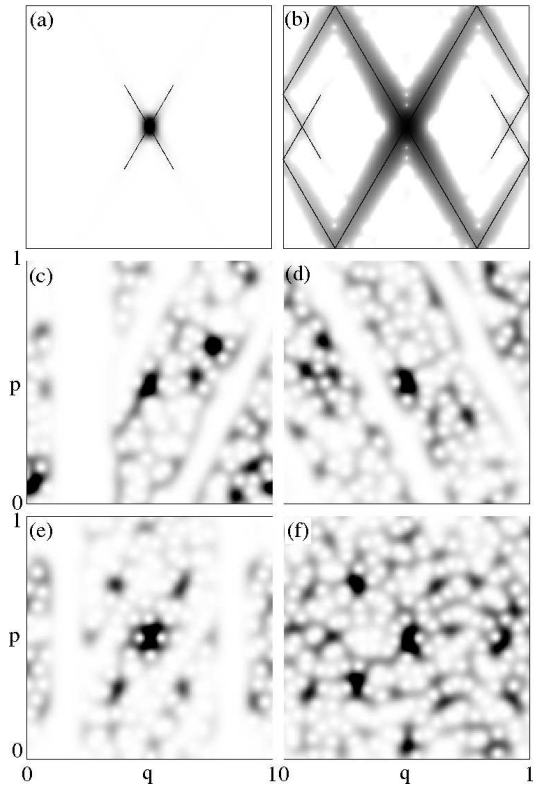}}
\caption{(a) Scar function of the PO given by $(q,p)=(0.5,0.5)$, 
for $N=225$. The horizontal axis corresponds to the 
position $q \in [0,1]$ and the vertical axis to the momentum 
$p \in [0,1]$ coordinate.
(b) Logarithmic version of (a). Black lines correspond 
to the stable and unstable manifolds of the orbit. 
(c) Right eigenvector of the ${\bar M_a}$ map which has the maximum 
overlap with the scar function shown in (a). (d) Same as previous 
panel but for the left eigenstate. (e) Same as (c) in the ${\bar M_b}$ 
map case. (f) Eigenfunction of the closed system 
having maximum overlap with the scar function of panel (a). 
See text for more details.}
\label{fig:scarcat}
\end{figure}

The main resource that we use to investigate localization is the scar 
function, which not only applies to maps but also to general flows. 
These functions have been deeply studied in the literature
\cite{ScarFunction1,ScarFunction2,ScarFunction3}. They are wavefunctions 
highly localized on the stable and unstable manifolds of POs, and 
on the energy given by a Bohr-Sommerfeld quantization 
condition on the trajectory.
We are going to use a formulation suitable for a Poincar\' e surface of 
section, or more directly for maps of the 2-torus 
(examples of this can be found in \cite{ScarFunction2}).
We define the ``tube functions'' for maps, $|\phi_{\rm tube}^{\rm maps}\rangle$ 
as a sum of coherent states 
centered at the fixed points of a given PO $\mu$, each one having 
a phase \cite{ScarFunction3}. Then, a dynamical average is performed, and we have the 
following expression for the scar function 
\begin{equation}
\label{eq:fscarm}
|\phi_{\rm scar}^{\rm maps}
\rangle= \sum_{l=-T}^T e^{i S_{\mu} l/\hbar} \; \cos \left(\frac{\pi l}{2 T} 
\right) \; M^l \; |\phi_{\rm tube}^{\rm maps}\rangle,
\end{equation}
where $T$ stands for the number of iterations of the map up to the Ehrenfest time
$T_E=\ln{\hbar}/\lambda$ ($\lambda$ is the Lyapunov exponent), and $S_{\mu}$ is the 
classical action of $\mu$.
We have used Eq. (\ref{eq:fscarm}) to construct functions highly 
localized on the vicinity of the periodic points of the closed cat map. 
In Fig. \ref{fig:scarcat}(a) we can see the structure of the scar function 
corresponding to the PO given by $(q,p)=(0.5,0.5)$, one of the shortest 
of this map, for $N=225$.
The maximum probabilities correspond to the darkest regions. Panel (b) of 
the same Figure displays this function in a logarithmic scale of gray, 
showing the way it extends along the stable and 
unstable manifolds of the corresponding orbit. 

\begin{figure}[htp] %[tp]
\centerline{\epsfxsize=7cm\epsffile{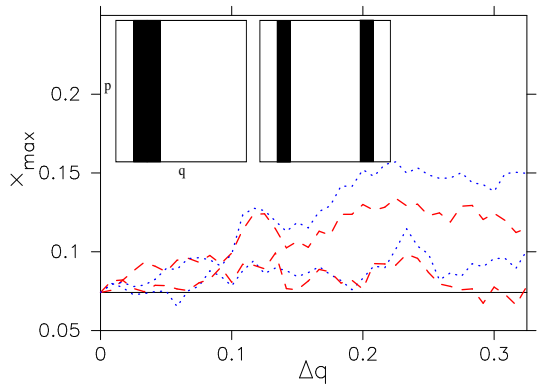}}
\caption{(Color online). Maximum overlap $x_{{\rm max}}$ (average from $N=350$ to $N=360$) 
of the scar function 
with the right (blue dotted) and left (red dashed) eigenstates of the open cat map 
as a function of the size of the opening $\Delta q$. The lower curves correspond 
to $\bar M_a$, while the upper ones to $\bar M_b$. The solid horizontal line 
stands for the value corresponding to the closed cat map. Left and right insets 
illustrate the projectors $P_a$ and $P_b$.}
\label{fig:maximaB}
\end{figure}

In the following we are going 
to describe the behaviour of localization in the open system by means of 
the maximum overlaps of the scar function with its resonances. We explore 
different values of $N$ and two different shapes of the projector $P$.
For simplicity we define a map ${\bar M_a=P_a M}$,  where $P_a$ 
corresponds to the projection on the complement of a band parallel to the 
$p$ direction of width $\Delta q$ and centered at $q=q_0$. 
We also define the map ${\bar M_b=P_b M}$, where we now consider two symmetric bands 
each one having a $\Delta q/2$ width, and centered at $q=q_0$ and $q=1-q_0$. 
This is shown in the left and right insets of Fig. \ref{fig:maximaB}.
In Fig. \ref{fig:scarcat}(c) and (d) we show the right 
and left eigenstates of the $\bar M_a$ map that have 
maximum overlap with the scar function displayed in Fig. \ref{fig:scarcat}(a) 
(here $q_0=0.225$ and $\Delta q=0.25$). The left resonance 
localizes on the unstable manifold, while the right one 
does it on the stable manifold.
The same can be found in panel (e) but for the $\bar M_b$ map 
($q_0=0.1625$). 
We can see that the symmetric cut localizes the resonance 
on the stable and unstable manifolds of the trajectory.  
Finally, in panel (f) the eigenfunction of the closed cat map 
with maximum overlap with the scar function of panel (a) is shown.

\begin{figure}[htp] %[tp]
\centerline{\epsfxsize=7cm\epsffile{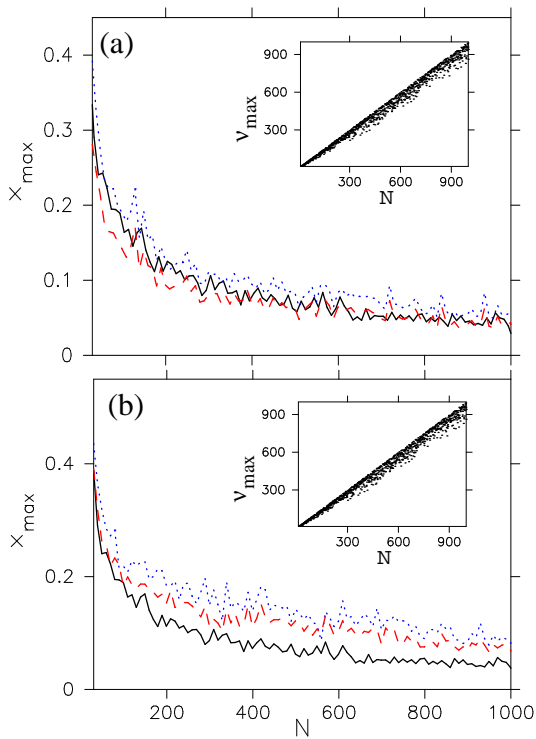}}
\caption{(Color online). Maximum overlap $x_{{\rm max}}$ of the scar function 
with the eigenstates of the open cat map as a function of $N$ 
(running average in a window $\Delta N=10$). 
In the insets we show the order number $\nu_{\rm max}$ in ascending eigenvalue 
modulus of the right resonance with maximum overlap, as a function of $N$ 
(for the left resonance, results are similar). 
Panel (a) corresponds to the ${\bar M_a}$ map, while the ${\bar M_b}$ map results 
are shown in (b). 
In all panels: the solid black lines correspond to the maximum overlap for the closed 
map, and the blue dotted and red dashed lines correspond to the maximum overlaps 
with the right and left resonances of the open maps, respectively.}
\label{fig:maximaA}
\end{figure}

First, we systematically analyze the behaviour of localization as a function 
of $\hbar$.
For that purpose, in Fig. \ref{fig:maximaA} we show the maximum 
overlaps of the scar function with 
the right and left eigenstates of the open cat map, as a function 
of $N$ (for clarity of the exposition  we show the running average of these 
values in a window of size $\Delta N=10$).  
It is evident that these values are greater for the 
open system, both for the right (blue dotted) and left (red dashed) resonances. 
On the other hand we can see the insets, where the order number $\nu_{\rm max}$ 
of the eigenstate with maximum overlap with the scar function is plotted 
vs. $N$ (they were ordered in ascending eigenvalue modulus). It is clear 
that the maximum overlap corresponds to resonances 
with the smaller decay rates (larger eigenvalue moduli). This guarantees that 
we are looking at wavefunctions which have a support on the classical repeller and 
do not belong to the null subspace. 
In Fig. \ref{fig:maximaA}(a) we can see the results for the ${\bar M_a}$ map, 
while in (b) they correspond to ${\bar M_b}$, where $q_0$ and $\Delta q$ values are 
taken the same as in the particular case of Fig. \ref{fig:scarcat}. 
The difference between the two maps turns out to be very important. In fact, the 
greater overlaps were obtained when opening in a symmetric fashion rather than with a single strip. 
Finally, we mention that the overlaps of the scar function with the left or the right resonances of 
the open map differ. These eigenstates are supported by different trapped classical sets, 
so in principle there is no reason for them to coincide. Anyway we think 
that the detailed explanation of this difference is an interesting 
open problem.

But then a natural question arises: how does the relationship between the shape and 
the size of the projection influences the intensity of scarring? For instance, 
this is relevant if we want to obtain highly localized resonances with the minimum 
amount of losses. This happens in many applications, the cases of two-dimensional 
billiards that can be used as optical microcavities for lasers or that can be 
attached to perfect leads, being some examples. 
To answer this we have further investigated 
the behaviour of localization by fixing the $\hbar$ value, and studying 
how the width of the opening influences it for both, $P_a$ and $P_b$ operators. 
The results are shown in 
Fig. \ref{fig:maximaB}, where we display the average of 
$x_{{\rm max}}$ taken from $N=350$ to $N=360$, as a function of 
the width of the opening $\Delta q$. In all cases we take $q_o=0.225$ for 
$\bar M_a$ and $q_o=0.1625$ for $\bar M_b$. 
The overlaps were calculated with the right (blue dotted) and left (red dashed) eigenstates. 
The lower curves correspond 
to $\bar M_a$, while the upper ones to $\bar M_b$. 
We have found that not only the overlap in general increases 
with the size of the opening, but also that this effect is greater 
due to the symmetrization. 

\begin{figure}[htp] %[tp]
\centerline{\epsfxsize=3.5cm\epsffile{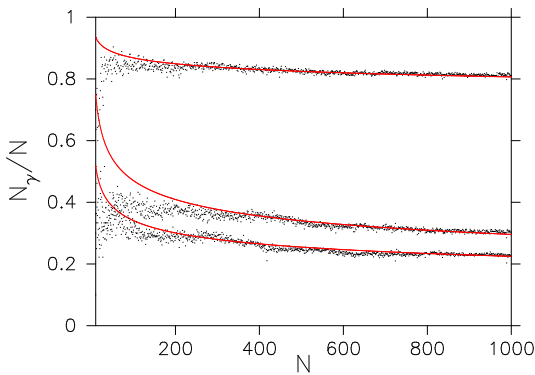}}
\caption{(Color online). Fraction of eigenstates $N_\gamma/N$ 
whose decay rate $\Gamma$ is smaller than $\Gamma_f=0.71$, 
as a function of $N$. The red lines corresponds to the 
theoretical expression $O^{(-1/(\lambda T_d))}$ (see 
text for details). The upper curve corresponds to ${\bar M_a}$ with an opening 
defined by $q_0=0.125$ and $\Delta q=0.05$, the middle one by $q_0=0.225$ 
and $\Delta q=0.25$, and finally the lower one corresponds to 
${\bar M_b}$ with $q_0=0.1625$ and $\Delta q=0.25$.}
\label{fig:WeylLaw}
\end{figure}

But this seemingly greater scarring effect in open systems should be interpreted 
in the proper context. 
In order to do this we will analyze the weight that these long-lived 
resonances have in the whole spectrum, and relate it with typical time scales of the system. 
This is given by a connection between the fractal Weyl law and the Ehrenfest time 
$T_o=\ln{(O)} / \lambda$ (with $O$ the number of open channels, and $\lambda=1.31$ in our 
case), first obtained in \cite{SchomerusPrl04}. There, it was found that the fraction of 
resonances with decay rate $\Gamma$ smaller than a fixed value $\Gamma_f<1/T_o$ 
behaves like $N_\gamma/N = \exp(-T_o/T_d)=O^{(-1/(\lambda T_d))}$, 
where $T_d=N/O$ is the so-called ``dwell time'' ($T_d$ large). 
We have numerically confirmed the validity of this prediction in our system by 
fitting the data with the expression $N_\gamma/N=a N^{-b}$. 
We show three cases in Fig. \ref{fig:WeylLaw}, where we have taken $\Gamma_f=0.71$ 
in all of them. 
The upper curve corresponds to ${\bar M_a}$ with an opening 
defined by $q_0=0.125$ and $\Delta q=0.05$, showing a fitted $b_f=0.032$ that 
agrees with the theoretical $b_{th}=0.038$. The middle curve corresponds to $q_0=0.225$ 
and $\Delta q=0.25$ with $b_f=0.181$, and the lower one 
corresponds to ${\bar M_b}$ with $q_0=0.1625$ and the same $\Delta q$  
with $b_f=0.2$, being $b_{th}=0.191$ for both.
In all cases this fraction goes to zero, leaving a small amount of 
classically meaningful eigenstates. 
This directly implies a persistent localization effect on the few remaining 
``fractal eigenfunctions''. However, these greater overlaps could correspond 
to a normalization difference with respect to the closed system. In fact, 
the effective size of the Hilbert space of the open 
system is smaller. Then, if we can go further and claim that 
these results mean a true enhancement of 
scarring, remains to be determined \cite{WisniackiFuture}. 

In summary, we have found that there is a greater overlap of the scar functions with 
the resonances of an open system compared to the closed one. 
The fractal structure of the eigenstates has been 
widely studied, motivating their denomination as ``quantum fractal 
eigenstates''. However, this significant alteration of the morphology 
of the eigenfunctions with respect to the analog closed system 
does not destroy the localization around POs. We think 
that this is due to the fact that the pruning of orbits that escape through 
the openings before the Ehrenfest time leaves parts of the stable and unstable 
manifolds. These remaining parts are enough to support the quantum structure 
associated to the scar function. However, the way they interfere in order 
to construct the same object than the smooth manifolds of the closed 
system remains unknown. Also, the scarring grows with the size 
of the opening ruling out any perturbative explanation for this. 
Moreover, this phenomenon is persistent, in the 
sense that it survives in the vanishing fraction of long-lived resonances 
as $N$ grows. 
In future studies \cite{WisniackiFuture} we will focus 
on these open questions.

%%%%%%%%%%%%%%%%%%%%%%%%%%%%%%%%%%%%%%%%%%%%%%%%%%%%%%%%%%%%%%%%%%%
\begin{acknowledgments}
The authors gratefully acknowledge support from CONICET (PIP-6137), 
UBACyT (X248), and ANPCyT. We have benefited from fruitful discussions 
with M. Saraceno and E. Vergini.
\end{acknowledgments}
%%%%%%%%%%%%%%%%%%%%%%%%%%%%%%%%%%%%%%%%%%%%%%%%%%%%%%%%%%%%%%%%%%%

%%%%%%%%%%%%%%%%%%%%%%%%%%%%%%%%%%%%%%%%%%%%%%%%%%%%%%%%%%%%%%%%%%%

\end{document}